\input harvmac
\input graphicx
\input color

\def\Title#1#2{\rightline{#1}\ifx\answ\bigans\nopagenumbers\pageno0\vskip1in
\else\pageno1\vskip.8in\fi \centerline{\titlefont #2}\vskip .5in}

%
%
\ifx\includegraphics\UnDeFiNeD\message{(NO graphicx.tex, FIGURES WILL BE IGNORED)}
\def\figin#1{\vskip2in}
\else\message{(FIGURES WILL BE INCLUDED)}\def\figin#1{#1}
\fi
\def\Fig#1{Fig.~\the\figno\xdef#1{Fig.~\the\figno}\global\advance\figno
 by1}
%
%
%
%

\font\ticp=cmcsc10

\def \purge#1 {\textcolor{magenta}{#1}}
\def \new#1 {\textcolor{blue}{#1}}
\def\comment#1{}

\def\\{\cr}
\def\text#1{{\rm #1}}
\def\frac#1#2{{#1\over#2}}

\def\hf{{1\over 2}} 
\def\calo{{\cal O}}

\def\roughly#1{\mathrel{\raise.3ex\hbox{$#1$\kern-.75em\lower1ex\hbox{$\sim$}}}}
\font\bbbi=msbm10 
\def\mathbb#1{\hbox{\bbbi #1}}

 \def\sgr{Sgr A${}^*$}

\def\mthsu{\mathsurround=0pt  }
\def\leftrightarrowfill{$\mthsu \mathord\leftarrow\mkern-6mu\cleaders
  \hbox{$\mkern-2mu \mathord- \mkern-2mu$}\hfill
  \mkern-6mu\mathord\rightarrow$}
\def\overleftrightarrow#1{\vbox{\ialign{##\crcr\leftrightarrowfill\crcr\noalign{\kern-1pt\nointerlineskip}$\hfil\displaystyle{#1}\hfil$\crcr}}}
\overfullrule=0pt

%
%
\lref\EHTr{The Event Horizon Telescope Collaboration, ``First M87 Event Horizon Telescope Results. I. The Shadow of the Supermassive Black Hole," 
ApJL {\bf 875} L1.}
\lref\AMPS{
  A.~Almheiri, D.~Marolf, J.~Polchinski and J.~Sully,
  ``Black Holes: Complementarity or Firewalls?,''
  JHEP {\bf 1302}, 062 (2013).
  [arXiv:1207.3123 [hep-th]].
}
\lref\MaSu{
  J.~Maldacena and L.~Susskind,
  ``Cool horizons for entangled black holes,''
Fortsch.\ Phys.\  {\bf 61}, 781 (2013).
[arXiv:1306.0533 [hep-th]].
}
\lref\NVNLT{
  S.~B.~Giddings,
 ``Modulated Hawking radiation and a nonviolent channel for information release,''
Phys.\ Lett.\ B {\bf 738}, 92 (2014).
[arXiv:1401.5804 [hep-th]].
}
\lref\BHMR{
  S.~B.~Giddings,
  ``Black holes and massive remnants,''
Phys.\ Rev.\ D {\bf 46}, 1347 (1992).
[hep-th/9203059].
}
\lref\BHQIUE{
  S.~B.~Giddings,
  ``Black holes, quantum information, and unitary evolution,''
  Phys.\ Rev.\ D {\bf 85}, 124063 (2012).
[arXiv:1201.1037 [hep-th]].
}
\lref\SGmodels{
  S.~B.~Giddings,
   ``Models for unitary black hole disintegration,''  Phys.\ Rev.\ D {\bf 85}, 044038 (2012)
[arXiv:1108.2015 [hep-th]].
}
\lref\NLvC{
  S.~B.~Giddings,
  ``Nonlocality versus complementarity: A Conservative approach to the information problem,''
Class.\ Quant.\ Grav.\  {\bf 28}, 025002 (2011).
[arXiv:0911.3395 [hep-th]].
}
\lref\NLEFTone{
  S.~B.~Giddings,
  ``Nonviolent information transfer from black holes: a field theory parameterization,''
Phys.\ Rev.\ D {\bf 88}, 024018 (2013).
[arXiv:1302.2613 [hep-th]].
}
\lref\NVNL{
  S.~B.~Giddings,
  ``Nonviolent nonlocality,''
  Phys.\ Rev.\ D {\bf 88},  064023 (2013).
[arXiv:1211.7070 [hep-th]].
}
\lref\Hawk{
  S.~W.~Hawking,
  ``Particle Creation By Black Holes,''
  Commun.\ Math.\ Phys.\  {\bf 43}, 199 (1975)
  [Erratum-ibid.\  {\bf 46}, 206 (1976)].
}
\lref\LPSTU{
  D.~A.~Lowe, J.~Polchinski, L.~Susskind, L.~Thorlacius and J.~Uglum,
  ``Black hole complementarity versus locality,''
Phys.\ Rev.\ D {\bf 52}, 6997 (1995).
[hep-th/9506138].
}
\lref\UnWamine{ W.~G.~Unruh and R.~M.~Wald
  ``How to mine energy from a black hole,''
Gen.\ Relat.\ Gravit.\ {\bf 15}, 195 (1983).}
\lref\stringmine{
 A.~E.~Lawrence and E.~J.~Martinec,
  ``Black hole evaporation along macroscopic strings,''
Phys.\ Rev.\ D {\bf 50}, 2680 (1994)
[hep-th/9312127]\semi
  V.~P.~Frolov and D.~Fursaev,
  ``Mining energy from a black hole by strings,''
Phys.\ Rev.\ D {\bf 63}, 124010 (2001)
[hep-th/0012260]\semi
  V.~P.~Frolov,
 ``Cosmic strings and energy mining from black holes,''
Int.\ J.\ Mod.\ Phys.\ A {\bf 17}, 2673 (2002).
}
\lref\LIGO{
  B.~P.~Abbott {\it et al.} [LIGO Scientific and Virgo Collaborations],
  ``Observation of Gravitational Waves from a Binary Black Hole Merger,''
Phys.\ Rev.\ Lett.\  {\bf 116}, no. 6, 061102 (2016).
[arXiv:1602.03837 [gr-qc]].
}
\lref\EHT{
  S.~Doeleman {\it et al.},
 ``Imaging an Event Horizon: submm-VLBI of a Super Massive Black Hole,''
[arXiv:0906.3899 [astro-ph.CO]].
}
\lref\Mathurrev{
  S.~D.~Mathur,
  ``Fuzzballs and the information paradox: A Summary and conjectures,''
[arXiv:0810.4525 [hep-th]].
}
\lref\Pageone{
  D.~N.~Page,
  ``Average entropy of a subsystem,''
Phys.\ Rev.\ Lett.\  {\bf 71}, 1291 (1993).
[gr-qc/9305007].
}
\lref\Pagetwo{
  D.~N.~Page,
  ``Information in black hole radiation,''
Phys.\ Rev.\ Lett.\  {\bf 71}, 3743 (1993).
[hep-th/9306083].
}
\lref\Brownmine{
  A.~R.~Brown,
  ``Tensile Strength and the Mining of Black Holes,''
Phys.\ Rev.\ Lett.\  {\bf 111}, no. 21, 211301 (2013).
[arXiv:1207.3342 [gr-qc]].
}
\lref\SGGW{
  S.~B.~Giddings,
  ``Gravitational wave tests of quantum modifications to black hole structure -- with post-GW150914 update,''
Class.\ Quant.\ Grav.\  {\bf 33}, no. 23, 235010 (2016).
[arXiv:1602.03622 [gr-qc]].
}
\lref\SGwind{
  S.~B.~Giddings,
  ``Possible observational windows for quantum effects from black holes,''
Phys.\ Rev.\ D {\bf 90}, no. 12, 124033 (2014).
[arXiv:1406.7001 [hep-th]].
}
\lref\SGnature{
  S.~B.~Giddings,
  ``Astronomical tests for quantum black hole structure,''
Nature Astronomy 1, Article number: 0067 (2017).
[arXiv:1703.03387 [gr-qc]].
}
\lref\NVU{
  S.~B.~Giddings,
  ``Nonviolent unitarization: basic postulates to soft quantum structure of black holes,''
JHEP {\bf 1712}, 047 (2017).
[arXiv:1701.08765 [hep-th]].
}
\lref\gravast{
  P.~O.~Mazur and E.~Mottola,
  ``Gravitational condensate stars: An alternative to black holes,''
[gr-qc/0109035].
}
\lref\SGobs{
  S.~B.~Giddings,
  ``Observational strong gravity and quantum black hole structure,''
Int.\ J.\ Mod.\ Phys.\ D {\bf 25}, no. 12, 1644014 (2016).
[arXiv:1605.05341 [gr-qc]].
}
\lref\GiPs{
  S.~B.~Giddings and D.~Psaltis,
  ``Event Horizon Telescope Observations as Probes for Quantum Structure of Astrophysical Black Holes,''
Phys.\ Rev.\ D {\bf 97}, no. 8, 084035 (2018).
[arXiv:1606.07814 [astro-ph.HE]].
}
\lref\GiRo{
  S.~B.~Giddings and M.~Rota,
  ``Quantum information or entanglement transfer between subsystems,''
Phys.\ Rev.\ A {\bf 98}, no. 6, 062329 (2018).
[arXiv:1710.00005 [quant-ph]].
}
\lref\LQGST{
  S.~B.~Giddings,
  ``Locality in quantum gravity and string theory,''
Phys.\ Rev.\ D {\bf 74}, 106006 (2006).
[hep-th/0604072].
}
\lref\BHIUN{
  S.~B.~Giddings,
 ``Black hole information, unitarity, and nonlocality,''
Phys.\ Rev.\ D {\bf 74}, 106005 (2006).
[hep-th/0605196].
}
\lref\DPPC{D. Psaltis, private communication.}
\Title{
\vbox{\baselineskip12pt  
}}
{\vbox{\centerline{Searching for quantum black hole structure} \centerline{with the Event Horizon Telescope
}}}

\centerline{{\ticp Steven B. Giddings\footnote{$^\ast$}{Email address: giddings@ucsb.edu}}}
\centerline{\sl Department of Physics}
\centerline{\sl University of California}
\centerline{\sl Santa Barbara, CA 93106}
\centerline{and}
\centerline{\sl CERN, Theory Department,}
\centerline{\sl 1 Esplande des Particules}
\centerline{\sl Geneva 23, CH-1211, Switzerland}

\vskip.10in
\centerline{\bf Abstract}
The impressive images from the Event Horizon Telescope sharpen the conflict between our observations of gravitational phenomena and the principles of quantum mechanics.  Two related scenarios for reconciling quantum mechanics with the existence of black hole-like objects, with ``minimal" departure from general relativity and local quantum field theory, have been explored; one of these could produce signatures visible to EHT observations.   A specific target is temporal variability of images, with a characteristic time scale determined by the classical black hole radius.  The absence of evidence for such variability in the initial observational span of seven days is not expected to strongly constrain such variability.  Theoretical and observational next steps towards investigating such scenarios are outlined.

\vskip.3in
\Date{}

The impressive success\refs{\EHTr} of the Event Horizon Telescope (EHT) in imaging of the apparent black hole in M87 has opened a second observational window to the regime of strong gravity, following the detection of gravitational waves from black hole mergers\refs{\LIGO}.  It has been widely believed that, due to the smallness of the curvature near the horizon of a large black hole (BH), this is a regime governed by classical general relativity (GR).  However, it has been increasingly appreciated that the inconsistency of BH evolution with the principles of quantum mechanics, given the Hawking effect\Hawk, strongly motivates the need to modify the classical description of a BH at {\it horizon scales}\refs{\BHMR\LQGST\BHIUN\Mathurrev\NLvC\AMPS-\MaSu}.  This suggests searching for signatures of such modification in  EHT or gravitational wave observations\refs{\SGwind\SGGW-\SGnature}.  As this note will elaborate on, a particularly interesting target is {\it temporal variability} of EHT images\refs{\GiPs}.

Without such modification of BH behavior on horizon scales, BH evolution appears to contradict quantum mechanics.  Suppose we think of a BH as a subsystem of a larger quantum system including its environment.  Either through interaction with surrounding matter and radiation, or just through the Hawking process itself, the BH subsystem will absorb information from its environment.\foot{More precisely, it develops entanglement with its environment.}  However, the Hawking process also indicates that BHs ultimately  evaporate away.  Locality of the geometric description of a BH states that a BH cannot release information while evaporating.  And, if an information-containing subsystem ceases to exist without releasing its information, that violates the basic quantum-mechanical principle of  unitary evolution.  

To preserve unitarity, BHs {\it must} release their information; simple arguments\refs{\Pageone,\Pagetwo} indicate that this should begin midway through a BH's decay, while the BH is comparable to its original size.  Such information release requires some modification to the description of a BH that is based local quantum field theory (LQFT) evolution on a classical BH geometry; this modification needs to occur on scales comparable to the size of the event horizon, for an arbitrarily large black hole.

In searching for the physics that makes BHs consistent with quantum mechanics, a reasonable approach is to seek a minimal departure from the conventional GR/LQFT description of BHs, possibly corrected for small backreaction effects.   In fact, such a ``correspondence principle" is a third postulate for BH evolution, beyond the postulates of consistency with quantum mechanics and of approximate validity of a subsystem description\refs{\NVU}.  

While there are various other conjectured scenarios for how BH evolution can be consistent with quantum mechanics\refs{\gravast,\Mathurrev,\AMPS}, the correspondence postulate strongly suggests an effective parameterization of BH evolution in terms of interactions between the BH quantum state, and the quantum degrees of freedom of the BH atmosphere.  Such an effective description has been developed in \refs{\NLvC,\SGmodels\BHQIUE-\NVNLT,\NVU}.

Beyond the three preceding postulates, a fourth\NVU\ is 
 that such interactions between a black hole and its environment couple universally to all degrees of freedom.  This is plausible for an effect that is intrinsically gravitational, and is also strongly motivated both by Gedanken experiments involving black hole mining\refs{\UnWamine,\stringmine} and by the desire to maintain as many features as possible of the beautiful story of BH thermodynamics. 

The interactions are most easily described in a hamiltonian approach, where they provide modifications of the usual hamiltonian evolution of LQFT, on a chosen set of time slices.\foot{For further details, see \refs{\NVU}.}  Combining the four postulates then strongly motivates interactions described by an effective Hamiltonian
\eqn\HInt{H_I = \int \sqrt{-g_{tt}}\,dV H^{\mu\nu}(x) T_{\mu\nu}(x)\ .}
Here $dV$ is the volume element on the time slices and $g_{tt}$ is the metric component in slice-time $t$, $T_{\mu\nu}(x)$ is the stress tensor for field excitations,\foot{This is also expected to contain perturbative gravitons.} and $H^{\mu\nu}(x)$ parameterize the interactions with the BH.\foot{In \HInt, $H_I$ is given in Schr\"odinger picture.}  Specifically, the $H^{\mu\nu}$ are operators that act on the BH quantum state, and these have spatial dependence that localizes them near the BH.  The simplest way to satisfy the correspondence principle is if these are localized within a distance of $\calo(R)$ from the event horizon of the classical geometry, of radius $R$, although more complicated scalings are possible\refs{\SGGW}.  One immediately sees that the interactions \HInt\ behave like {\it BH state dependent metric perturbations}.  And this raises the prospect of observational signatures in EHT images, due to the effect of such interactions on light propagation near the horizon.

A key question for prospects of observing  such a ``quantum halo" is the size of the  $H^{\mu\nu}$.  In units, $\hbar=c=1$, $H^{\mu\nu}$ is dimensionless, as is appropriate for a metric perturbation.  Since $H^{\mu\nu}$ is an operator on BH states, one characteristic of its size is its  expectation value $\langle H^{\mu\nu}\rangle$ in a typical BH state.
Another key question is the time dependence of $H^{\mu\nu}$.  In a Schr\"odinger picture description, $H^{\mu\nu}$ can be time independent, but from the viewpoint of the BH exterior it gains time dependence because it mediates transitions between BH quantum states.  Again, by correspondence, the simplest possibility is if $H^{\mu\nu}$ dominantly connects states with energy separations $\Delta E\sim 1/R$, and thus $H^{\mu\nu}$ produces exterior excitations with energies comparable to that of Hawking quanta.

The size of  $H^{\mu\nu}$ is constrained by the requirement of unitary evolution.  Arguments related to those of \refs{\Pageone,\Pagetwo} indicate that the interaction \HInt\ needs to transfer information from the BH at a rate
\eqn\inforate{{dI\over dt}\sim {1\over R}\ ,}
{\it i.e.}, at roughly one qubit per light crossing time.  A bilinear hamiltonian of the form \HInt\ will generically transfer information, but one needs to determine the rate at what it does so.

The simplest way to achieve the required rate \inforate\ follows purely from dimensional analysis.  All important dimensional quantities have scale governed by $R$, and so this rate arises from
\eqn\Hcoh{\langle H^{\mu\nu}(x)\rangle\sim 1\ ,}
for $x$ an $\calo(R)$ distance from the horizon.
Such an $\calo(1)$ metric perturbation raises the prospect of important observational consequences.

One can inquire whether \Hcoh\ is the {\it minimum} possible size.  This leads to general question in information theory\refs{\NVU,\GiRo}: given two subsystems coupled by a Hamiltonian with a bilinear interaction like \HInt, how fast does information transfer?  In the BH context, ref.~\NVU\ argued that the operators $H^{\mu\nu}$ could have a much smaller typical size than \Hcoh\ and still achieve the rate \inforate, due to a quantum enhancement in the rate if there is a large number of final possible states.  BHs are expected to have an enormous number of possible internal states, 
\eqn\BHstates{N\sim e^{S_{bh}}\ ,}
where, for example $S_{bh}$ might be given by the Bekenstein-Hawking entropy, Area/$4G$.  
A simple argument for the rate estimate follows from Fermi's Golden Rule, and yields the result that 
\eqn\Hincoh{\langle H^{\mu\nu}\rangle\sim 1/{\sqrt N}\ }
can achieve the rate \inforate.

Thus, there are two distinct possible scenarios, the {\it strong/coherent} scenario of \Hcoh, and the {\it weak/incoherent} scenario of \Hincoh.  

Clearly for an astronomical-sized black hole, \Hincoh\ leads to tiny metric perturbations, which might be na\"\i vely expected to have small effects.  However, consider scattering from a BH, mediated by the interaction \HInt.  Fermi's Golden Rule can be again used to estimate the transition probability, which takes the form
\eqn\scatprob{P \sim  \vert \langle \alpha |H_I|\beta\rangle\vert^2\rho(E)\ ,}
where the states $\alpha$, $\beta$ include the initial and final states of the scattered particle, and $\rho$ is the density of final states.  By the same logic as above, this can be $\calo(1)$, given a typical size \Hincoh.  However, while the transition probability can be $\calo(1)$, this occurs for transitions with momentum transfer $\sim1/R$.  For a $mm$ wavelength photon scattering off \sgr, the fractional momentum transfer is $\calo(10^{-13})$, which is negligible.  However, for gravitational waves, which in binary mergers commonly have wavelength $\sim R$, this is potentially a relevant effect.  Since the present discussion is of EHT observations, without prejudicing which answer Nature prefers we will focus attention on the strong/coherent scenario for the remainder of this note.

To reiterate the relevant parameters for the simplest version of a strong/coherent scenario \Hcoh, it specifies that there are $\calo(1)$ perturbations in the effective metric, spatially varying over a scale $\sim R$ outside what would be the horizon, and time-varying on scales $\sim R$. This suggests more detailed examination of observational prospects.

A preliminary investigation of possible impact of such perturbations on BH images was conducted in \refs{\GiPs}, based on ray tracing in the presence of simple examples for such perturbations.  These simple examples demonstrated significant distortion of the simulated images, compared to those from an unperturbed black hole.  This work thus serves as an important check that sensitivity to time-varying perturbations of size \Hcoh\ is clearly possible.

The initially-reported EHT observations\EHTr\ spanned a seven day period; within instrumental resolution, there was no clear indication for temporal variability.\foot{Apparent variations in reconstructed images can so far be attributed to observational/instrumental systematic effects\refs{\DPPC}.}  
In order to test the strong/coherent scenario, it is obviously important to go beyond these initial impressive results, with both further theoretical and observational work.

On the theoretical side, one problem is to more completely explore the parameter space of possible perturbations, and the dependence of observable signatures on these parameters.  Ref.~\GiPs\ considered a simple parameterization of such coherent perturbations, in terms of their radial wavenumber $k$, their radial extension $R_G$ above the horizon, their angular profile determined by spherical harmonics with angular momenta $l,m$, their amplitude, and their frequency $\omega$.  The example spectrum of perturbations assumed in \GiPs\ produced a rather dramatic time dependence of images.  These include changes in the shape and size of the shadow and surrounding ring.  However, variation of these parameters can of course yield much less dramatic effects.  It is also important to extend the models of \GiPs\ to the case of a black hole with angular momentum.

In particular, the fact that the initial EHT images produced a clearly identifiable  shadow should place an important bound on this parameter space, which could be inferred with further work to study dependence of the simulated images on the  parameters.

A particularly promising signature is the temporal variability, which has prospects to be a key test for the effects described here.  A first question is the relevant time scale, particularly given the initially limited observation window.  The characteristic time scale that was used in the simulations of \GiPs\ is determined by the constraint that the perturbations do minimal damage to the thermodynamic description of BHs.  
Specifically, if there are interactions of the form \HInt, where $H^{\mu\nu}$ effectively varies with frequency $\omega$, that will create radiation from the BH with frequency $\sim \omega$.  So, the minimal departure from the thermal Hawking spectrum arises if $\omega \approx \omega_T$, given by the Hawking temperature, where for a BH with mass $M$ and angular momentum $L=a GM^2$, 
\eqn\Therm{\omega_T = {1\over 4\pi GM}\, {\sqrt{1-a^2}\over 1+\sqrt{1-a^2}}\ .}
The corresponding periods are 
\eqn\Periods{\eqalign{
 P= \frac{2\pi}{\omega_{\rm T}}  
  &\simeq 0.93 \left(\frac{M}{4.3\times 10^6 M_\odot}\right)\left(\hf + {1\over 2\sqrt{1-a^2}}\right)~\mbox{hr}\cr 
  &\simeq  59 \left(\frac{M}{6.5\times 10^9 M_\odot}\right)\left(\hf + {1\over 2\sqrt{1-a^2}}\right)~\mbox{d}\;, }}
for \sgr\ and M87, with respective masses $M\simeq   4.3\times 10^6 M_\odot$ and $6.5\times 10^9 M_\odot$. 

For \sgr, such a period is short as compared to the multi-hour imaging scan time used by EHT, except in case of high spin,\foot{Note that a spin parameter $a=.98$ multiplies the period by 3.0.} and so EHT images will average over fluctuations on these time scales.  The EHT images are not produced from a simple average, but rather based on a more complex algorithm involving the magnitudes of Fourier components and closure phases of baseline triangles.  This presents a problem requiring more analysis, to determine the impact of the perturbations on these averaged images, once they are obtained, and their distinguishability from other sources of variability, such as turbulence in the accretion.

For M87, it appears easier to see such a  time dependence directly\GiPs.  However, the currently-reported observations span a time of seven days.  While relaxing the demand for agreement with BH thermodynamics could yield shorter-time scale variations, investigating the scale \Periods\ requires longer-span observations. If the spin is high, this period may be significantly extended.  Thus, on the observational end, longer-span observations will be useful in searching for strong/coherent fluctuations.

It is also important to disentangle any observed variability from other possible origins; in addition to turbulence in the accretion flow, there can be scattering from the interstellar medium.  This is another place where combined theoretical and observational efforts may be needed, both in order to better model such time-dependence, particularly from a conventional accretion origin, and to investigate ways to better distinguish these possible origins observationally.  One important aspect of this\GiPs\ is the achromaticity of effective metric perturbations; they affect light of all frequencies uniformly, 
so that their effects should be observed at different EHT operating wavelengths.  

Further such theoretical and observational investigations are important for determining EHT sensitivity to the strong/coherent scenario.
Ultimately such efforts could reveal the existence of strong/coherent perturbations, or rule them out over a large range of parameter space.  The latter would point towards the weak/incoherent scenario, or to an even more exotic reconciliation of the existence of black-hole like objects with quantum mechanics.

\bigskip\bigskip\centerline{{\bf Acknowledgments}}\nobreak

I thank  the CERN theory group, where this work was carried out, for its hospitality.
This material is based in part upon work supported in part by the U.S. Department of Energy, Office of Science, under Award Number {DE-SC}0011702.

\listrefs
\end